# Curiosity as a Self-Supervised Method to Improve Exploration in De novo Drug Design

Mohamed-Amine CHADI
Computer science department
Faculté des sciences Semlalia
Université Cadi Ayyad
Marrakech, Morocco
mohamedamine.chadi@ced.uca.ma

Hajar Mousannif
Computer science department
Faculté des sciences Semlalia
Université Cadi Ayyad
Marrakech, Morocco
mousannif@uca.ac.ma

Ahmed Aamouche
LISA laboratory
National school of applied sciences
Cadi Ayyad university
Marrakech, Morocco
a.aamouche@uca.ma

*Abstract*—In recent years, deep learning has demonstrated promising results in de novo drug design. However, the proposed techniques still lack an efficient exploration of the large chemical space. Most of these methods explore a small fragment of the chemical space of known drugs, if the desired molecules were not found, the process ends. In this work, we introduce a curiosity-driven method to force the model to navigate many parts of the chemical space, therefore, achieving higher desirability and diversity as well. At first, we train a recurrent neural network-based general molecular generator (G), then we fine-tune G to maximize curiosity and desirability. We define curiosity as the Tanimoto similarity between two generated molecules, a first molecule generated by G, and a second one generated by a copy of G (Gcopy). We only backpropagate the loss through G while keeping Gcopy unchanged. We benchmarked our approach against two desirable chemical properties related to drug-likeness and showed that the discovered chemical space can be significantly expanded, thus, discovering a higher number of desirable molecules with more diversity and potentially easier to synthesize. All Code and data used in this paper are available at https://github.com/amine179/Curiosity-RL-for-Drug-Design.

*Keywords—Drug design; Curiosity; reinforcement learning; Chemical space exploration*

## I. INTRODUCTION

The chemical space of potential drug-like molecules is huge with a known estimation of ~$10^{60}$ compounds in size [1]. Consequently, the process of de novo drug design and discovery is easily prone to failure despite the relatively high-cost and long working time put into it [2]. In this context, several deep learning-based techniques have been proposed in the last five years as a promising alternative to traditional virtual screening-based methods [3].

Often, deep learning-based approaches consist of training a deep generative model (e.g., recurrent neural network (RNN) [4]) to learn the underlying rules and structural patterns of chemically valid molecules, then, applying a fine-tuning technique such as transfer learning (TL) [5] or reinforcement learning (RL) [6] to focus the generative model on a minute part of the chemical space that is hopefully rich in terms of desired drug candidates. However, the exploration capacity of the proposed techniques is relatively limited and poorly addressed.

In this paper, the concept of chemical space refers to a cluster of all molecules with a defined quantifiable structural or chemical similarity.

Over the last years, deep learning has emerged as a powerful tool in the field of computational de novo drug design. A relatively popular one is the work by [7]. Here, the authors utilized deep RL to bias an RNN-molecular generator to design molecules with optimized properties such as lipophilicity (logP) and the half-maximal inhibitory concentration ($IC_{50}$) against the Janus protein kinase 2. This same pipeline was used in other research to generate molecules for different targets, such as dopamine receptor type 2 in [8] and adenosine $A_{2A}$ receptor in [9], as well as other chemical and structural *per se* optimization objectives as in [10] and [11] respectively.

Besides, when there is a dataset available for the task at hand, but in a small quantity, transfer learning comes in handy. For this, other researchers exploited TL as the main technique to bias the general generative model. For instance, [12] fine-tuned an RNN-based generative model using a set of molecules belonging to the space of SARS-CoV- $M^{pro}$ inhibitors. In [13], the authors were able to reproduce 14% of 6051 and 28% of 1240 already available molecules designed by medicinal chemists. Finally, [14] and [15] demonstrated the usefulness of TL in this context against other general use-case chemical objectives.

Despite the efforts, the presented approaches have so far focused solely on the desirability, that is, identifying some part of the chemical space with a potentially high number of desirable molecules. However, to maximize the chance of discovering potential hits, the diversity of the space should also be considered during the optimization process. This is of great importance because, although the model may generate drugs with optimized chemical properties, they might be synthetically infeasible in practice. Moreover, current techniques do not ensure that the learned part of the chemical space is the "richest" in terms of desirable molecules, therefore, exploiting multiple parts of the space can be of huge benefit.

In this work, we propose a curiosity-driven RL approach to explore multiple promising parts of the chemical space. Curiosity-driven RL was popularized by [16] and refers to an intrinsic signal (i.e., coming from the agent itself) that quantifies how certain the agent is in terms of predicting the future consequences of its actions. By seeking the minimization of the accuracy of this prediction, the agent is said to become curious in the sense that it strives to explore other unseen areas. It was used as a replacement for the extrinsic reward (i.e., reward coming from the environment) and other traditional reward augmentation methods such as reward shaping [17], and has demonstrated better performance than several state-of-the-art algorithms, especially in complex exploration tasks. In our approach, we define curiosity in such a way as to quantify the difference between the molecule generated by the generative model (G) and a molecule generated by a copy of the same generative model (Gcopy). By forcing G to maximize this difference, we incentivize it to change its focus from a previously learned part of the chemical space to another one. More details about this pipeline are presented in the next section (Methods).

## II. METHODS

### A. Data and the generative model

To train our generative model, we used the CHEMBL21 dataset [18] which contains ~1.6 million drug-like molecules represented as strings known as the simplified molecular line-entry system (SMILES) format. In the SMILES format, atoms are depicted using corresponding letters, and special characters such as (e.g., =, #, @, etc.) denote bonding types as well as the opening and closure of rings and branches. Accordingly, we chose a type of neural network architecture that is suitable for such sequential data, which is the RNN model, specifically, RNN with the gated recurrent units (GRU) cells [19]. Recurrent neural networks are designed to utilize the patterns learned across the sequential steps by learning to memorize the most important information of past steps for the prediction of the current ones.

To this end, we formulate the problem of molecular generation as a sequential character-by-character classification, that is, given the $i^{th}$ character of a molecule (in the SMILES format), what is the probability of each character in the vocabulary to be the $i^{th+1}$ character (see Fig. 1, step 1). To concretize such output, we use the Softmax distribution (described in Equation 1) at the end of our model, with a temperature parameter T allowing for further tuning.

$$P(yp) = \frac{e^{(yp/T)}}{\sum_{j=1}^{k} e^{(yp/T)}} \quad (1)$$

The higher T (>1) is, the more the probability of all characters is equal, while lower T (<1) values increase the probability of characters of already high probabilities (i.e., in the mean region) and increase it for others.

To evaluate the learning of the model, a loss function should be defined. For this, we use the cross entropy (CE) loss defined in Equation 2 which maximizes the (log) likelihood of selecting the right characters, which in turn results in generating molecules with similar patterns to the training data.

$$CE(yp, yr) = -\frac{1}{k} \sum_{i=1}^{k} yr * log(yp) \quad (2)$$

Where yp and yr are the predicted and actual output, respectively, and k is the number of classes (characters).

The architecture of the generative model, referred to as G in the remainder of the paper, is illustrated in the architecture is composed of an embedding layer, three GRU layers of 512 units, and a final linear layer. The output of the linear layer is then fed into the Softmax layer that uses Equation 1 to output a distribution probability of all characters which we sample from a predicted output. We trained G for 2500 epochs, with a learning rate of 0.0005 using the Adam optimizer [20].

### B. Reinforcement learning

Reinforcement learning (RL) [6] involves learning a policy function, which is a mapping of states (input) to actions (output) to maximize a numerical reward signal. The generative RNN model only mimics the training data, but with no specific constraints or objectives. To increase the probability of molecules that satisfy a certain objective/chemical property (e.g., logP <=3), RL can be utilized by assigning higher rewards to the generated molecules with the desired property and lower rewards for others (illustrated in Fig. 1, step 2). In the remainder of this paper, we refer to this focused model as RL-G-0.

For the specific RL algorithm, we use the REINFORCE algorithm [21]. REINFORCE is a policy gradient algorithm, meaning that the policy parameters are updated according to the direct estimate of the gradient of the expected sum of rewards (Equation 3), as opposed to other RL methods that rely on some indirect estimates such as the action-value (Q-value) or the state-value (V) [6]. This makes it easy to integrate with architectures such as the RNN model used in this work.

$$J(\theta) = E[r(s, a)] = \sum_{t=0}^{T} p(s_t, a_t)(r_t) \rightarrow max \mid s_0, \theta \quad (3)$$

Where θ is the set of parameters of the policy (generative model), E is the mathematical expectation, r is the reward, s is the state (the input character), and a is the action (the output character).

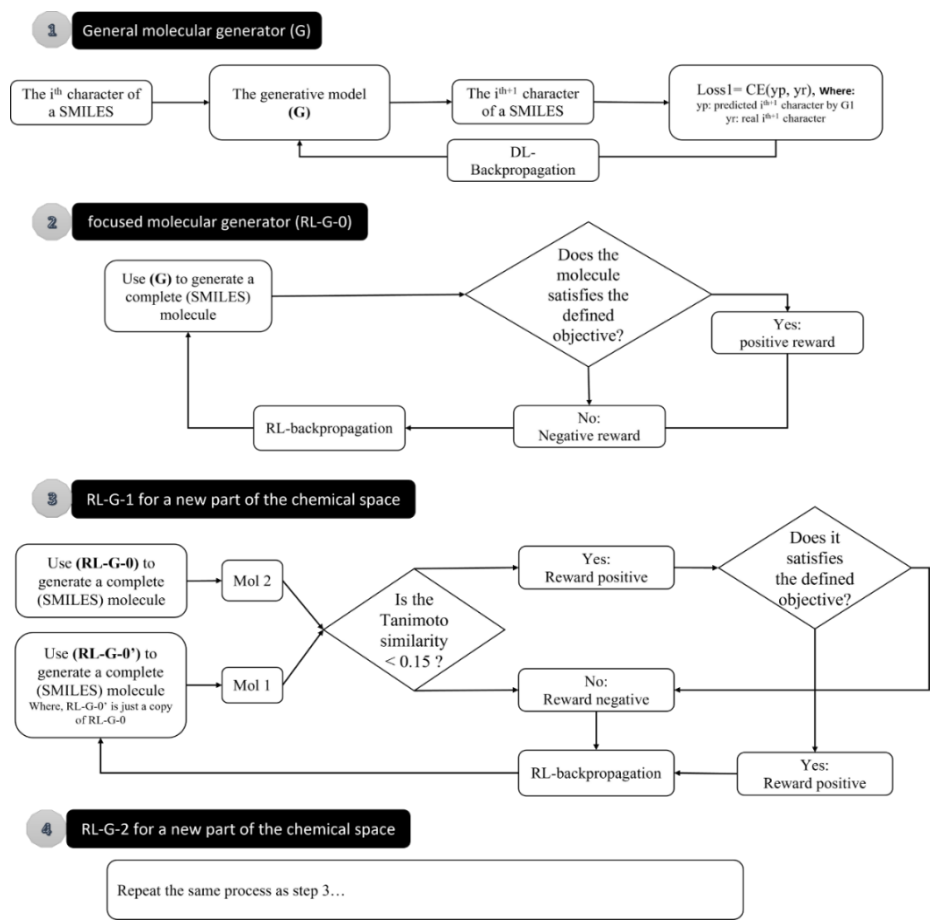

Fig. 1 the global pipeline of the proposed approach: (1) training the general model, (2) focusing it using reinforcement learning, and (3-4) changing its focus to other parts of the chemical space.

### C. Integrating curiosity

Once the RL-G-0 model was trained, we will be able to use it to generate desirable molecules. However, these molecules are prone to low diversity, especially for highly constrained objectives. For this, it is necessary to search for other promising parts of the large chemical space. To force our model to change its focus towards other promising spaces, we propose to integrate the curiosity mechanism and obtain a new version of the model, named RL-G-1 (or RL-G-i, where i ≥ 1). We suggest achieving this by fine-tuning RL-G-0 to, not only generate desirable molecules but also, to decrease the Tanimoto similarity [22] between the molecule generated by RL-G-0 and its successor RL-G-1 (generalized to RL-G-i and RL-G-i+1) as described in Fig. 1, step 3. By forcing the model to generate desirable and structurally novel molecules (with respect to the previous version of the model), the model is said to become curious, in the sense that it seeks to explore a new space of solutions in a new part of the chemical space.

### D. Objectives (target molecules)

To evaluate the developed approach, we defined two distinct types of objectives (single and multi-objective). The multi-objective is the famous Lipinski rule of three [23] (a modified version of the Lipinski rule of five), which states that orally bioavailable compounds are likely to have the following properties:
- logP ≤ 3,
- molecular weight ≤ 480 g/mol,
- hydrogen bond acceptors ≤3,
- hydrogen bond donors ≤3,
- rotational bonds ≤3.

To this end, the single objective we set is the first component of the Lipinski rule, that is, generated molecules should have a logP ≤ 3. While the multi-objective is the full Lipinski rule, that is, generated molecules should satisfy all five components. All chemical properties were computed using the freely available chemistry library RDKit [24].

### E. Evaluation metrics

For each version of the focused generative models (RL-G-0, RL-G-1, RL-G-2, etc.), five metrics were computed:
- Validity: the proportion of valid molecules, computed using RDKit.

- Novelty: the proportion of novel molecules (i.e., not present in the CHEMBL dataset) among valid ones.
- Uniqueness: the proportion of molecules without duplicates among novel ones.
- Internal diversity (intDiv): computed using the open-source library MOSES [25]. It assesses the chemical diversity within the generated valid, novel, and unique molecules using the Tanimoto similarity index. This metric ranges between 0 and 1, with values close to 1 denoting high diversity.
- Desirability: the proportion of valid, novel, and unique molecules with the desired properties.

### III. RESULTS AND DISCUSSION

After many trials for setting a value for the temperature, T=0.50 yielded optimal results of 81%±1.6, 91%±2.1, 89%±1.5, and 0.81±0.98 for validity, novelty, uniqueness, and internal diversity, respectively.

To understand the core idea of this work, we shall go over the results presented in Table 1 along with Fig. 2. While Table 1 presents the benchmarking results in terms of the metrics discussed previously, Fig. 2 illustrates the uniform manifold approximation and projection (UMAP) [26] plots of Morgan circular fingerprints [27] of length 1,024 bits and radius of 2 bonds. The purpose of Fig. 2 is to track the change of the chemical space focused on by a specific version of the model.

In Table 1 we show the benchmarking results of our biased model (RL-G-0) as well as four runs of the curiosity-driven models RL-G-1 to RL-G-4. We see that for both objectives (single and multi), the first biased model (RL-G-0) has successfully generated desirable molecules with a good diversity value of 0.81. Moreover, we observe that for the single objective, the learned chemical space overlaps moderately with that of the training data that satisfy the same objective. For the multi-objective RL-G-0, the learned chemical space is completely new compared to that of the training data. However, to make sure that this is the optimal part of the chemical space (i.e., the one with the highest number of desirable molecules) we should uncover the other parts as well, and this is where curiosity-driven models fit in. In the first curiosity-based model (RL-G-1) for the single objective, we show that indeed the model switched its focus from the previously learned chemical space to another one, thus demonstrating the effectiveness of our approach. Nevertheless, the number of desirable molecules generated was reduced, which suggests that this part of the chemical space is a sub-optimal one. As for the multi-objective version of RL-G-1, and because curiosity incentivizes the model to avoid only the space of the previous model, we observe that this one tended to reconstruct molecules that belong to the space of the training data, which might be useful for some machine learning applications such as data augmentation. After repeating this process three other times, we had in our hands one normal RL-biased model (RL-G-0) and four curiosity-RL-biased models (RL-G-1 to RL-G-4). Each one of the curiosity-based models has shown a similar pattern of performance: a degradation in the number of desirable generated molecules, and the covering of a new part of the chemical space.

We believe this approach can find many interesting applications in the field of deep learning-based molecular generation such as drug design as it allows for a broader range of different molecules to investigate. Indeed, although the chemical space uncovered by the first biased model (RL-G-0) by the mean of RL might be richer in terms of desirability, it does not ensure easy-to-synthesize drugs. To demonstrate this, we present in Fig. 3 the kernel density estimation (KDE) plots of the synthetic accessibility score (SAS) [28] of the generated molecules by each model, respectively, computed using the MOSES library [25]. SAS reflects on a scale of 0 to 10, how difficult it is to synthesize a given molecule, where molecules with higher values of SAS, typically above 6, are considered difficult to synthesize in practice. Results illustrated in Fig. 3 show that it could be a benefit in exploring other parts of the chemical space. For instance, the chemical spaces covered by RL-G-2, RL-G-3, and RL-G-4 are richer in terms of molecules with low SAS values than previous ones, as well as the one covered by the general model G.

TABLE I. BENCHMARKING RESULTS

| Model | Objective | Validity | Novelty | Uniqueness | intDiv | Desirability |
|---|---|---|---|---|---|---|
| RL-G-0 | Single | 80.96% | 80.17% | 77.44% | 0.81 | 39.56% = **1989** |
| RL-G-0 | Multi | 86.73% | 55.22% | 51.44% | 0.80 | 27.75% = 684 |
| RL-G-1 | Single | 67.64% | 61.94% | 75.05% | 0.78 | 30.77% = **968** |
| RL-G-1 | Multi | 89.85% | 48.90% | 39.25% | 0.77 | 20.17% = 348 |
| RL-G-2 | Single | 55.02% | 54.12% | 69.40% | 0.81 | 42.18% = **872** |
| RL-G-2 | Multi | 89.65% | 54.23% | 42.45% | 0.76 | 18.07% = 373 |
| RL-G-3 | Single | 85.63% | 32.51% | 45.79% | 0.81 | 41.72% = **532** |
| RL-G-3 | Multi | 82.33% | 40.27% | 31.00% | 0.79 | 28.50% = 293 |
| RL-G-4 | Single | 86.48% | 37.23% | 39.75% | 0.82 | 43.28% = **554** |
| RL-G-4 | Multi | 85.40% | 48.45% | 27.45% | 0.78 | 24.29% = 276 |

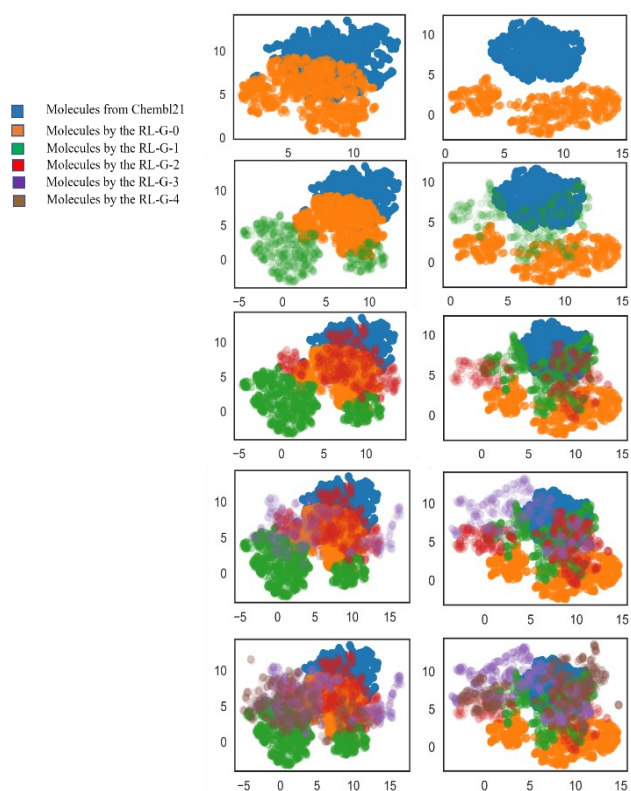

Fig. 2 UMAP visualization of the chemical spaces of desirable molecules from several sources

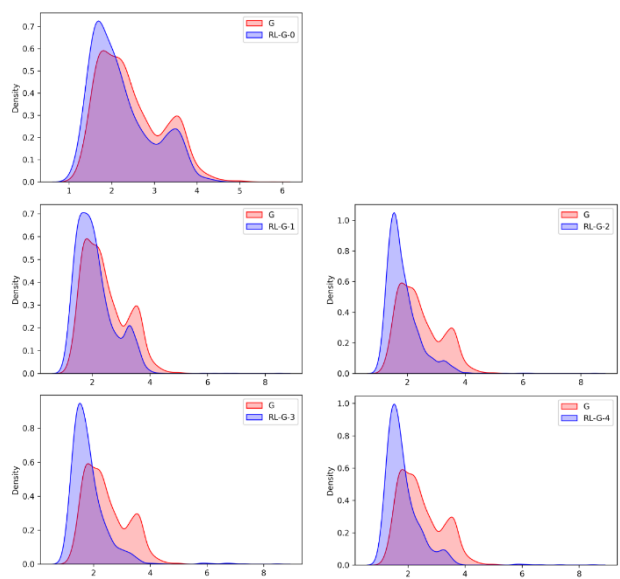

Fig. 3 SAS distributions of the generated molecules by each biased model compared to that of G.

## IV. CONCLUSION

The importance of this approach is to cover the largest chemical space of potential desirable molecules to better ensure optimality and provide practitioners of chemical engineering with more diverse sets of molecules for in-lab tests. In the above-presented experiments and results, we demonstrated the usefulness of curiosity-driven RL in the context of drug design as it allows for successively changing the focus of a deep learning-based molecular generator. As far as our knowledge, this is the first study of this kind, therefore, our goal was to provide researchers with a starting point and benchmarking results to compare with future improvement addressing this same important issue.